\documentclass[english, aps,prd,twocolumn,twoside,superscriptaddress,floatfix,nofootinbib]{revtex4-1}

\usepackage[utf8]{inputenc}
\usepackage{amsmath}
\usepackage{geometry}
\usepackage[dvipsnames]{xcolor}
\usepackage{mathrsfs}
\usepackage{graphicx}
\usepackage{cancel}
\usepackage{float}
\usepackage{lipsum, babel}
\usepackage{mathtools,leftindex,tensor,mhchem}
\usepackage{tensor}
\usepackage{graphicx}
\usepackage{color}
\usepackage[colorlinks=false]{hyperref}

\usepackage{ifthen,amsmath,amssymb,bm}

\newcommand{\beq} {\begin{equation}}
\newcommand{\eeq} {\end{equation}}
\newcommand{\bal} {\begin{aligned}}
\newcommand{\eal} {\end{aligned}}

\begin{document}

\title{Bias hardened estimators of patchy screening profiles}

\author{Noah Sailer}
\email{nsailer@berkeley.edu}
\affiliation{Berkeley Center for Cosmological Physics, Department of Physics,
University of California, Berkeley, CA 94720, USA}
\affiliation{Physics Division, Lawrence Berkeley National Laboratory, Berkeley, CA 94720, USA}

\author{Boryana Hadzhiyska}
\affiliation{Miller Institute for Basic Research in Science, University of California, Berkeley, CA, 94720, USA}
\affiliation{Physics Division, Lawrence Berkeley National Laboratory, Berkeley, CA 94720, USA}
\affiliation{Berkeley Center for Cosmological Physics, Department of Physics,
University of California, Berkeley, CA 94720, USA}

\author{Simone Ferraro}
\affiliation{Physics Division, Lawrence Berkeley National Laboratory, Berkeley, CA 94720, USA}
\affiliation{Berkeley Center for Cosmological Physics, Department of Physics,
University of California, Berkeley, CA 94720, USA}

\begin{abstract}
Detecting anisotropic screening of the cosmic microwave background (CMB) holds the promise of revealing the distribution of gas in the Universe, characterizing the complex processes of galaxy formation and feedback, and studying the epoch of reionization. Estimators for inhomogeneous screening, including some recently proposed small-scale (stacked) estimators, are quadratic or higher order in the CMB temperature or polarization fields and are therefore subject to contamination from CMB lensing. We review the origin of this lensing bias and show that, when stacking on \textit{unWISE} galaxies, the expected lensing bias dominates the signal if left unmitigated. Hardening techniques that null the lensing bias have been proposed for standard quadratic estimators, whereas only approximate methods have been proposed for stacked estimators. We review these techniques and apply the former to stacked estimators, presenting several strategies (including the optimal strategy) to null lensing contamination when stacking on any large-scale structure (LSS) tracer.
\end{abstract}

\maketitle

\section{Introduction}
Studying the diffuse gas in the Universe is of paramount importance to characterize the complex processes behind galaxy formation and feedback. Moreover, the ionized gas amounts to almost 16\% of the mass in the Universe, making understanding its location essential for the cosmological interpretation of weak lensing measurements from current and future surveys. High resolution and low noise observations of the cosmic microwave background (CMB) make it possible to use the CMB as a ``backlight'' to study the gas at low redshift. Several effects imprint the properties of the gas on the small-scale CMB. In particular, two of the Sunyaev-Zel'dovich (SZ) effects \cite{1981ASPRv...1....1S} are proportional to the gas density: the kinematic SZ (kSZ) effect is the Doppler shift of a photon's energy due to a galaxy's peculiar velocity, and is proportional to both the gas density and the large scale peculiar velocity. The patchy screening (or ``blurring'' SZ) effect is the suppression of the CMB anisotropies by Thomson scattering with free electrons, and is proportional to both the gas density and the large-scale CMB temperature or polarization fluctuations. While kSZ has been used to measure the gas properties for quite some time \cite{ACTPol:2015teu, AtacamaCosmologyTelescope:2020wtv, ACT:2024vsj}, patchy screening has been less studied in the literature: \cite{Dvorkin:2008tf} showed that quadratic estimators similar to those of CMB lensing can be used to reconstruct the optical depth $\tau$ field, while \cite{2011arXiv1106.4313S} showed that CMB lensing can be an important contribution and suggested a method of lens-hardening to mitigate that problem\footnote{Ref.~\cite{2011arXiv1106.4313S} also derives a lensing estimator that is unbiased to patchy screening.}. Later forecasts show that the $\tau$ field can be used to characterize the epoch of reionization \cite{2013PhRvD..87d7303G} and the gas around late-time galaxies \cite{Roy:2022muv}. 

Following similar work for estimators of the kSZ effect \cite{Li:2014mja, ACTPol:2015teu} and cluster lensing \cite{Horowitz:2017iql}, stacking estimators for $\tau$ from patchy screening were proposed by refs.~\cite{Schutt:2024zxe,ACT:2024rue}. These stacked estimators are quadratic (or higher order) in the underlying CMB temperature field, and therefore are sensitive to the effect of CMB lensing.
In this paper, we explain the origin of the lensing contamination to stacked estimators and discuss three possible mitigation strategies with varying levels of assumptions made about the lensing field.

The remainder of the paper is organized as follows: in \S\ref{sec:patchy_screening_estimators} we briefly review several patchy screening estimators. In \S\ref{sec:lens_hardened_stacked_estimators} we present three approaches to harden these estimators against lensing, which we validate using simulations in \S\ref{sec:sims}. 
We comment on the importance of lensing biases for \textit{unWISE} galaxies in \S\ref{sec:unwise} and conclude in \S\ref{sec:discussion_and_conclusion}. In a companion paper \cite{boryana_companion_paper} we present a self-consistent comparison of kSZ- and patchy screening-inferred electron profiles taking into account the lensing bias explored here.

\textit{Notation:} Since we primarily work on small scales, we adopt the flat-sky approximation throughout. We let $\bm{r}$ and $\bm{\ell}$ correspond to angular positions and momenta respectively. We adopt the following convention for Fourier transforms
\begin{equation*}
    T_{\bm{\ell}} 
    =
    \int d^2\bm{r}\,
    e^{-i\bm{\ell}\cdot\bm{r}}\,
    T(\bm{r})
    \quad
    {\rm and}
    \quad
    T(\bm{r})
    =
    \int_{\bm{\ell}}
    e^{i\bm{\ell}\cdot\bm{r}}\,
    T_{\bm{\ell}}
\end{equation*}
where $\int_{\bm{\ell}} = \int d^2\bm{\ell}/(2\pi)^2$. Throughout we let $\langle \cdots \rangle$ denote a full ensemble average and $\langle \cdots\rangle'$ an average over primary CMB realizations in the presence of a fixed late-time large scale structure (LSS) realization.


\section{Patchy screening estimators}
\label{sec:patchy_screening_estimators}

Thomson scattering of CMB photons in and out of the line of sight damps primary CMB temperature fluctuations according to:
\begin{equation}
\label{eq:patchy_screening}
    T(\bm{r}) 
    =
    e^{-\tau(\bm{r})} T_0(\bm{r})
    \simeq T_0(\bm{r}) - \tau(\bm{r}) T_0(\bm{r}),
\end{equation}
where $T$ is the observed CMB temperature fluctuation, $T_0$ is the primary CMB and $\tau\ll1$ is the optical depth field. 
The modulation of the primary CMB by the optical depth produces (among other things) a non-zero $\langle TT\tau\rangle$ bispectrum.
The $\tau$-induced bispectrum can equivalently be phrased in terms of a \textit{linear response} function: when averaging over primary CMB realizations in the presence of a \textit{fixed} $\tau$, the covariance $\langle T_{\bm{\ell}} T_{\bm{L}-\bm{\ell}}\rangle'$ acquires ``off diagonal" (or anisotropic) contributions proportional to the optical depth \cite{Dvorkin:2008tf}:
\begin{equation}
\begin{aligned}
\label{eq:patchy_tau_response}
    &\langle T_{\bm{\ell}} T_{\bm{L}-\bm{\ell}}\rangle'
    =
    (2\pi)^2\delta^D_{\bm{L}} C^{TT}_\ell
    +
    f^\tau_{\bm{\ell},\bm{L}-\bm{\ell}}
    \,\tau_{\bm{L}}
    +
    \mathcal{O}(\tau^2)
    \\
    &{\rm with}\quad
    f^\tau_{\bm{\ell},\bm{L}-\bm{\ell}}
    =
    -\Big[C^{TT}_\ell + C^{TT}_{|\bm{L}-\bm{\ell}|}\Big],
\end{aligned}
\end{equation}
where $f^\tau$ is the \textit{linear response} of the CMB to patchy screening. To relate $f^\tau$ to the bispectrum, multiply both sides of Eq.~\eqref{eq:patchy_tau_response} by $\tau_{-\bm{L}}$ and perform a full ensemble average, yielding $f^\tau_{\bm{\ell},\bm{L}-\bm{\ell}} = \langle T_{\bm{\ell}}T_{\bm{L}-\bm{\ell}} \tau_{-\bm{L}} \rangle/\langle \tau_{\bm{L}}\tau_{-\bm{L}} \rangle$.

\subsection{Optimal quadratic estimator}

To reconstruct patchy screening from the observed CMB maps ref.~\cite{Dvorkin:2008tf} proposed a quadratic estimator (QE) $\hat{\tau}$, which takes the general form
\begin{equation}
\label{eq:general_qe}
    \hat{\tau}_{\bm{L}}
    =
    \int_{\bm{\ell}}
    F^\tau_{\bm{\ell},\bm{L}-\bm{\ell}}
    T_{\bm{\ell}}
    T_{\bm{L}-\bm{\ell}},
\end{equation}
for a set of weights $F^\tau_{\bm{\ell},\bm{L}-\bm{\ell}}$. To ensure that the estimator has unit response to patchy screening (i.e. $\langle \hat{\tau}\rangle' = \tau$) we enforce the normalization constraint $\int_{\bm{\ell}} F^\tau_{\bm{\ell},\bm{L}-\bm{\ell}} f^\tau_{\bm{\ell},\bm{L}-\bm{\ell}}=1$. In particular, the minimum variance weights satisfying the normalization constraint are given by \cite{2002ApJ...574..566H}
\begin{equation}
\label{eq:min_var_qe}
    F^{\tau,{\rm MV}}_{\bm{\ell},\bm{L}-\bm{\ell}}
    =
    \frac{f^\tau_{\bm{\ell},\bm{L}-\bm{\ell}}}{2 C^{\rm tot}_\ell C^{\rm tot}_{|\bm{L}-\bm{\ell}|}}
    \Bigg/
    \int_{\bm{\ell}}
    \frac{\big(f^\tau_{\bm{\ell},\bm{L}-\bm{\ell}}\big)^2}{2 C^{\rm tot}_\ell C^{\rm tot}_{|\bm{L}-\bm{\ell}|}},
\end{equation}
where $C^{\rm tot}_\ell$ is the total observed temperature power spectrum.

\subsection{Stacked estimators}

Since the ionized gas surrounding galaxies has a $\mathcal{O}(1)$ arcmin extent, the optical depth field is primarily composed of small scale modes. On the other hand, Silk damping largely erases these modes from the primary CMB. Motivated by this insight, ref.~\cite{Schutt:2024zxe} proposed using a long-short split to estimate the optical depth field. To do so we construct low- and high-pass filtered CMB maps with isotropic filters $W^l_\ell$ and $W^s_\ell$ respectively, which we colloquially refer to as long and short filters. 
The long map $T^l_{\bm{\ell}} \equiv W^l_\ell T_{\bm{\ell}} \simeq T^0_{\bm{\ell}}$ is predominately sensitive to the primary CMB, while the short map $T^s_{\bm{\ell}} \equiv W^s_\ell T_{\bm{\ell}} \simeq -\int_{\bm{\ell}'} \tau_{\bm{\ell}'} T^0_{\bm{\ell}-\bm{\ell}'}$ is predominantly sensitive to the modulation of the primary CMB by patchy screening. 
Ref.~\cite{Schutt:2024zxe} uses this to construct an optical depth map estimator from the ratio $-T_s(\bm{r}) / T_l(\bm{r})$ which under idealistic conditions is significantly less noisy than the optimal QE on small scales.

Alternatively, we can use the long-short split to construct a real space QE: $-T_l(\bm{r}) T_s(\bm{r}) / \langle T_l^2\rangle$. While the real space QE is suboptimal (i.e. its weights differ from Eq.~\ref{eq:min_var_qe}) it has the advantage of being computationally simpler than a full sky QE formalism, is readily amenable to public stacking codes\footnote{e.g. \url{https://github.com/EmmanuelSchaan/ThumbStack}} \cite{AtacamaCosmologyTelescope:2020wtv} and can be easily adapted to perform oriented stacking \citep[following e.g.][]{2012ApJ...758...74B,Planck:2015igc,ACT:2021iyk,ACT:2024fqd}. When stacking the real space QE on a set of galaxy positions $\bm{r}_i$ we obtain a \textit{real space stacked estimator} $\widehat{\mathcal{T}}$ for the optical depth profile, which can be (schematically) interpreted as the average optical depth at a displacement $\bm{r}$ away from a galaxy in the sample:
\begin{equation}
\label{eq:stacked_estimator}
\boxed{
    \widehat{\mathcal{T}}(\bm{r})
    =
    \frac{-1}{\langle T_l^2\rangle N_g}
    \sum_{i=1}^{N_g}
    T_l(\xi\bm{r}+\bm{r}_i) T_s(\bm{r}+\bm{r}_i), 
}
\end{equation}
where we introduced a number $\xi$ that specifies the offset between where $T_l$ and $T_s$ are evaluated. When comparing with simulations in \S\ref{sec:sims} we consider only two cases: $\xi=0$ or $1$, corresponding to the long map being evaluated at the galaxy center or a distance $r$ away from the galaxy center respectively. When $r$ is smaller than the scale associated with the long-short split $T_l(\bm{r})$ is roughly a constant 
and the estimator's response to patchy screening only weakly varies with $\mathcal{O}(1)$ changes in $\xi$. We explicitly show this in \S\ref{sec:stacked_hardening}. 

\section{Lens hardened stacked estimators}
\label{sec:lens_hardened_stacked_estimators}

Weak gravitational lensing of the CMB effectively shifts CMB pixels from position $\bm{r}$ to $\bm{r}+\bm{\nabla}\phi(\bm{r})$, where $\phi$ is the lensing potential \cite{2006PhR...429....1L}. To linear order this shift is equivalent to adding the term $\bm{\nabla}\phi\cdot\bm{\nabla}T_0$ to Eq.~\eqref{eq:patchy_screening}. Following the same logic as \S\ref{sec:patchy_screening_estimators}, the lensing term induces its own off-diagonal contributions to the covariance $\langle T_{\bm{\ell}}T_{\bm{L}-\bm{\ell}}\rangle' = f^\tau_{\bm{\ell},\bm{L}-\bm{\ell}}\tau_{\bm{L}}+f^\kappa_{\bm{\ell},\bm{L}-\bm{\ell}}\kappa_{\bm{L}}+\cdots$, where $\kappa_{\bm{L}} = L^2\phi_{\bm{L}}/L^2$ is the CMB lensing convergence and 
\begin{equation}
    f^\kappa_{\bm{\ell},\bm{L}-\bm{\ell}}
    =
    \frac{2\bm{L}}{L^2}
    \cdot
    \Big[
    \bm{\ell} C^{TT}_\ell
    +
    (\bm{L}-\bm{\ell}) C^{TT}_{|\bm{L}-\bm{\ell}|}
    \Big]
\end{equation}
is the associated linear response to lensing \cite{2002ApJ...574..566H}. Since the stacked estimator is quadratic (or higher order, see Appendix \ref{sec:general_stacked_linear_response}) in the CMB temperature it is subject to biases from CMB lensing \cite{2011arXiv1106.4313S}. Here we explore three strategies to mitigate CMB lensing biases to stacked patchy screening estimators with varying levels of assumptions made about the lensing field. 

\subsection{Field level hardening}
\label{sec:field_level_hardening}
Rather than the real space stacked estimator (Eq.~\ref{eq:stacked_estimator}), one could in principle use the minimum variance QE (Eqs.~\ref{eq:general_qe} \& \ref{eq:min_var_qe}) to estimate the optical depth profile by averaging $\hat{\tau}^{\rm MV}(\bm{r}+\bm{r}_i) = \int_{\bm{L}} e^{i\bm{L}\cdot(\bm{r}+\bm{r}_i)}\hat{\tau}^{\rm MV}_{\bm{L}}$ over galaxy positions. Following the discussion above, the minimum variance patchy screening estimator acquires a bias from lensing $\langle\hat{\tau}^{\rm MV}_{\bm{L}} \rangle' = \tau_{\bm{L}} + R^{\tau\kappa}_L\kappa_{\bm{L}}$, where we introduced the response function
\begin{equation}
    R^{\tau\kappa}_L
    =
    \int_{\bm{\ell}}
    F^{\tau,{\rm MV}}_{\bm{\ell},\bm{L}-\bm{\ell}}
    f^{\kappa}_{\bm{\ell},\bm{L}-\bm{\ell}}.
\end{equation}

To remove this bias ref.~\cite{2011arXiv1106.4313S} constructs a minimum variance CMB lensing estimator $\hat{\kappa}^{\rm MV}$, which can be obtained by relabeling $\tau\to\kappa$ in Eqs.~\eqref{eq:general_qe} and \eqref{eq:min_var_qe}. The minimum variance lensing estimator has a bias proportional to the optical depth: $\langle\hat{\kappa}^{\rm MV}_{\bm{L}} \rangle' = \kappa_{\bm{L}} + R^{\kappa\tau}_L\tau_{\bm{L}}$. By taking an appropriate linear combination of the minimum variance $\tau$ and $\kappa$ estimators, ref.~\cite{2011arXiv1106.4313S} showed that one can subtract off the lensing bias while retaining unit response to the optical depth 
\citep[see also][]{2013MNRAS.431..609N,Namikawa:2013xka,2014JCAP...03..024O,Sailer:2020lal,Namikawa:2021zhh}. Alternatively one can obtain a bias hardened $\tau$ estimator by imposing $\int_{\bm{\ell}}F^\tau_{\bm{\ell},\bm{L}-\bm{\ell}} f^\kappa_{\bm{\ell},\bm{L}-\bm{\ell}} =0$ in addition to the normalization constraint and performing a constrained minimization \cite{2023PhRvD.107b3504S}. Both approaches yield the same result for the optimal lens-hardened weights:
\begin{equation}
\label{eq:bias_hardened_weights}
\boxed{
\begin{aligned}
    F^{\tau,{\rm BH}}_{\bm{\ell},\bm{L}-\bm{\ell}}
    &=
    \frac{
    F^{\tau,{\rm MV}}_{\bm{\ell},\bm{L}-\bm{\ell}} - 
    R^{\tau\kappa}_L F^{\kappa,{\rm MV}}_{\bm{\ell},\bm{L}-\bm{\ell}}
    }{
    1 - R^{\tau\kappa}_L R^{\kappa\tau}_L
    }.
\end{aligned}
}
\end{equation}
With these modified weights, the average of $\hat{\tau}^{\rm BH}(\bm{r}+\bm{r}_i)$ over galaxy positions is an unbiased estimate of the optical depth profile.

\begin{figure}[!h]
    \centering
    \includegraphics[width=\linewidth]{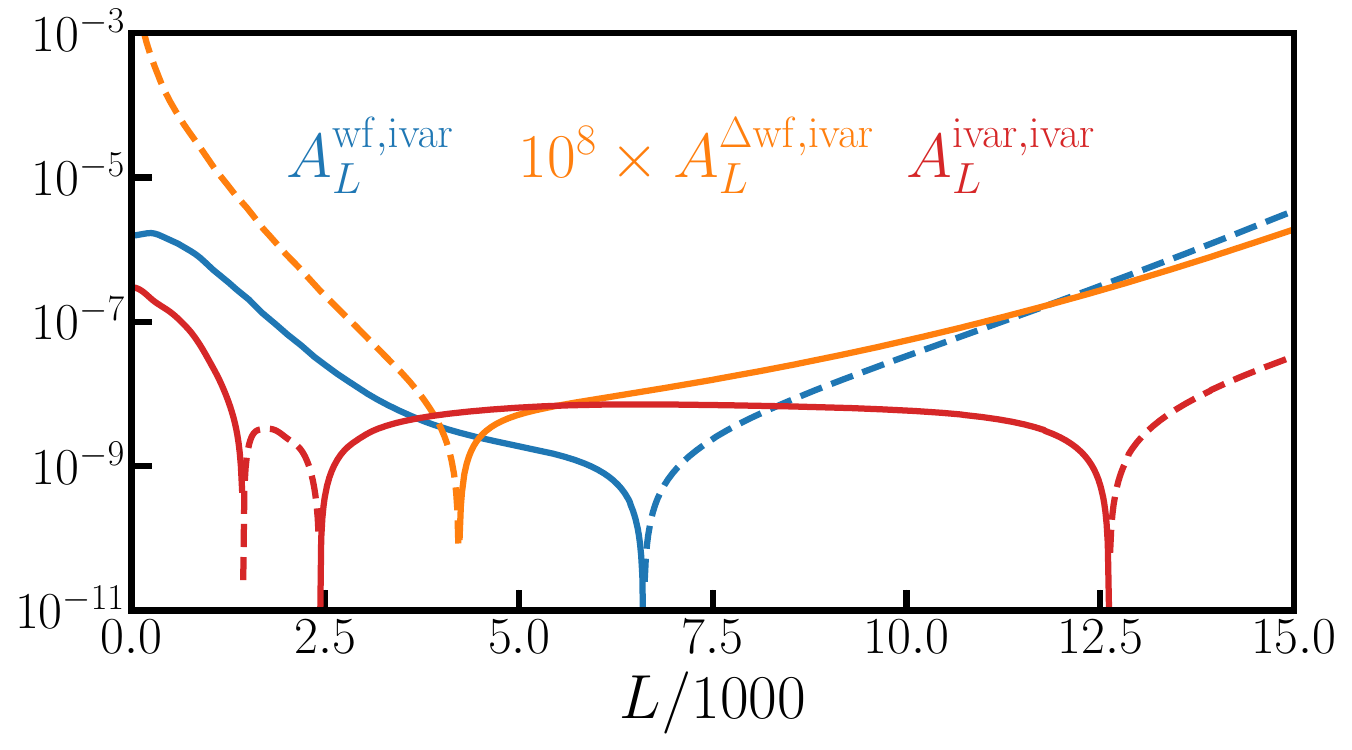}
    \caption{The three coefficients appearing in Eq.~\eqref{eq:kappa_ps_hardening} for the experimental setup discussed in \S\ref{sec:sims}, with the color labeling the coefficient. Solid (dashed) lines correspond to positive (negative) values.}
    \label{fig:bias_hardened_filters}
\end{figure}

It is straightforward to generalize this procedure to simultaneously harden against both lensing and unclustered point sources \cite{Namikawa:2021zhh}. A tedious but straightforward calculation shows that the resulting bias hardened QE $\hat{\tau}^{\rm BH}$ can be expressed as a series of filtering and Fourier transforms: 
\begin{widetext}
\begin{equation}
\label{eq:kappa_ps_hardening}
    \hat{\tau}^{\rm BH}_{\bm{L}}
    =
    A^{\rm wf, ivar}_L
    \mathcal{F}_{\bm{L}}
    \Big[
    T^{\rm wf} \times T^{\rm ivar}
    \Big]
    +
    A^{\Delta{\rm wf, ivar}}_L
    \mathcal{F}_{\bm{L}}
    \Big[
    T^{\rm wf} \times \big(\nabla^2T\big)^{\rm ivar}
    -
    \big(\nabla^2 T\big)^{\rm wf} \times T^{\rm ivar}
    \Big]
    +
    A^{\rm ivar,ivar}_L
    \mathcal{F}_{\bm{L}}
    \Big[
    \big(T^{\rm ivar}\big)^2
    \Big],
\end{equation}
where $T^{\rm wf}_{\bm{\ell}} = C^{TT}_\ell\,T_{\bm{\ell}}/C^{\rm tot}_\ell$ and $T^{\rm ivar}_{\bm{\ell}} = T_{\bm{\ell}}/C^{\rm tot}_\ell$ are Wiener- and inverse variance-filtered maps respectively, $(\nabla^2 T)_{\bm{\ell}} = - \ell^2 T_{\bm{\ell}}$ is a gradient map and $\mathcal{F}_{\bm{L}}[\cdots]$ indicates a Fourier transform of a real-space map. The coefficients $A^{\rm wf, ivar}$, $A^{\Delta {\rm wf, ivar}}$ and $A^{\rm ivar, ivar}$ (see Fig.~\ref{fig:bias_hardened_filters}) are given by
\begin{equation}
\begin{aligned}
    &\frac{A^{\rm wf,ivar}_L}{N^{\tau,{\rm BH}}_L}
    =
    \frac{r^{\tau s}_L r^{\kappa s}_L - (r^{\kappa s}_L)^2}{r^{\kappa\kappa}_Lr^{ss}_L}
    -
    \frac{r^{\tau\kappa}_L}{r^{\kappa\kappa}_L}
    -1
    ,\quad
    \frac{A^{\Delta{\rm wf,ivar}}_L}{N^{\tau,{\rm BH}}_L}
    =
    \frac{1}{L^2}
    \bigg[
    \frac{r^{\tau s}_L r^{\kappa s}_L}{r^{\kappa\kappa}_Lr^{ss}_L}
    -
    \frac{r^{\tau\kappa}_L}{r^{\kappa\kappa}_L}
    \bigg]
    ,\quad
    \frac{A^{{\rm ivar,ivar}}_L}{N^{\tau,{\rm BH}}_L}
    =
    \frac{r^{\tau \kappa}_L r^{\kappa s}_L}{r^{\kappa\kappa}_Lr^{ss}_L}
    -
    \frac{r^{\tau s}_L}{r^{ss}_L},
    \\
    &{\rm where}\quad
    \Big(N^{\tau,{\rm BH}}_L\Big)^{-1}
    =
    r^{\tau\tau}_L
    \bigg[
    1
    - \frac{(r^{\tau\kappa}_L)^2}{r^{\tau\tau}_L r^{\kappa\kappa}_L}
    - \frac{(r^{\tau s}_L)^2}{r^{\tau\tau}_L r^{ss}_L}
    - \frac{(r^{\kappa s}_L)^2}{r^{\kappa\kappa}_L r^{ss}_L}
    + 2\frac{r^{\tau \kappa}_Lr^{\tau s}_L r^{\kappa s}_L}{r^{\tau\tau}_L r^{\kappa\kappa}_Lr^{ss}_L}
    \bigg]
    \quad
    {\rm and}
    \quad
    r^{XY}_L
    =
    \int_{\bm{\ell}}
    \frac{
    f^X_{\bm{\ell},\bm{L}-\bm{\ell}} 
    f^Y_{\bm{\ell},\bm{L}-\bm{\ell}} 
    }{
    2 C^{\rm tot}_\ell C^{\rm tot}_{|\bm{L}-\bm{\ell}|}
    }.
\end{aligned}
\end{equation}
\end{widetext}
Here the superscript ``$s$" refers to point sources, for which the corresponding response function is scale invariant (i.e. $f^{s}_{\bm{\ell},\bm{L}-\bm{\ell}} =1$).
The quantity $r^{XY}_L$ can be thought of as the unnormalized response of the MV $X$ estimator to $Y$, while the properly normalized response is given by $R^{XY}_L = r^{XY}_L/ r^{XX}_L$. 
Note that the simplified scenario of only hardening against lensing (Eq.~\ref{eq:bias_hardened_weights}) can be obtained by setting $r^{\tau s}=r^{\kappa s}=0$ in the equation above. 


\subsection{Stacked hardening}
\label{sec:stacked_hardening}

Field level hardening (\S\ref{sec:field_level_hardening}) removes the leading order lensing bias via an internal quadratic estimate of the lensing convergence. On the small scales relevant for patchy screening measurements going beyond the QE will be more optimal and accurate for both estimating the patchy screening signal and removing any lensing contamination \citep[see e.g.][]{2007NJPh....9..441H,Horowitz:2017iql,Millea:2021had,Bianchini:2022wte,2024JCAP...01..024S}.
While in this work we restrict ourselves to ``stacked estimates" of patchy screening (e.g. Eq.~\ref{eq:stacked_estimator}), in this section we will assume that we are given a general \textit{unbiased estimate} $\hat{\kappa}(\bm{r})$ of the convergence field which could come from e.g. a QE, a maximum likelihood approach, or external LSS data as a lensing template.
Below we present an analogous method to lens harden the real space stacked estimator $\hat{\mathcal{T}}$ (Eq.~\ref{eq:stacked_estimator}) using this unbiased estimate of the convergence field, which we call \textit{stacked hardening}.

To connect stacked hardening to the previous section we simply take $R_L^{\kappa\tau}\to0$ in Eq.~\eqref{eq:bias_hardened_weights}. That is, the lensing estimator is presumed to be unbiased to patchy screening.
In this limit there is a simple interpretation of how one estimates the lensing bias to the stacked minimum variance QE: filter the convergence estimate by the response function $R^{\tau\kappa}_L$, transform back to real space and stack on the galaxy positions.

We generalize this procedure to the real space stacked estimator in Appendix \ref{sec:general_stacked_linear_response}, where we compute the linear response $\mathcal{R}^{\mathcal{T}\kappa}_{\bm{L}}(\bm{r})$ of the stacked estimator to the lensing convergence. Following the same recipe as before we obtain an estimator for the lensing bias by filtering the convergence estimator and averaging over galaxy positions:
\begin{equation}
\label{eq:stacked_hardening}
\boxed{
\begin{aligned}
    &{{\rm lensing\,\,bias}}\,(\bm{r})
    =
    \frac{1}{N_g}
    \sum_{i=1}^{N_g}
    \hat{\kappa}^F(\bm{r}_i,\bm{r}),
    \\
    &\hat{\kappa}^F(\bm{r}_i,\bm{r})
    =
    \int_{\bm{L}}
    e^{i\bm{L}\cdot\bm{r}_i}
    {\mathcal{R}}^{\mathcal{T}\kappa}_{\bm{L}}(\bm{r})
    \hat{\kappa}_{\bm{L}},
    \\
    &{\mathcal{R}}^{\mathcal{T}\kappa}_{\bm{L}}(\bm{r})
    =
    \frac{-1}{\langle T_l^2\rangle}
    \int_{\bm{\ell}}
    e^{i\big(\bm{L}-(1-\xi)\bm{\ell}\big)\cdot\bm{r}}
    W^s_{|\bm{L}-\bm{\ell}|} W^l_{\ell}
    f^\kappa_{\bm{\ell},\bm{L}-\bm{\ell}}.
\end{aligned}
}
\end{equation}
Note that this procedure can be straightforwardly adapted to oriented stacking by taking $\hat{\kappa}^F(\bm{r}_i,\bm{r})\to \hat{\kappa}^F(\bm{r}_i,\mathsf{R}_i\,\bm{r})$, where $\mathsf{R}_i$ is a rotation matrix for the $i$'th galaxy. Additionally, one can easily generalize this approach to more optimally weight the galaxy sample.

\begin{figure}[!h]
    \centering
    \includegraphics[width=\linewidth]{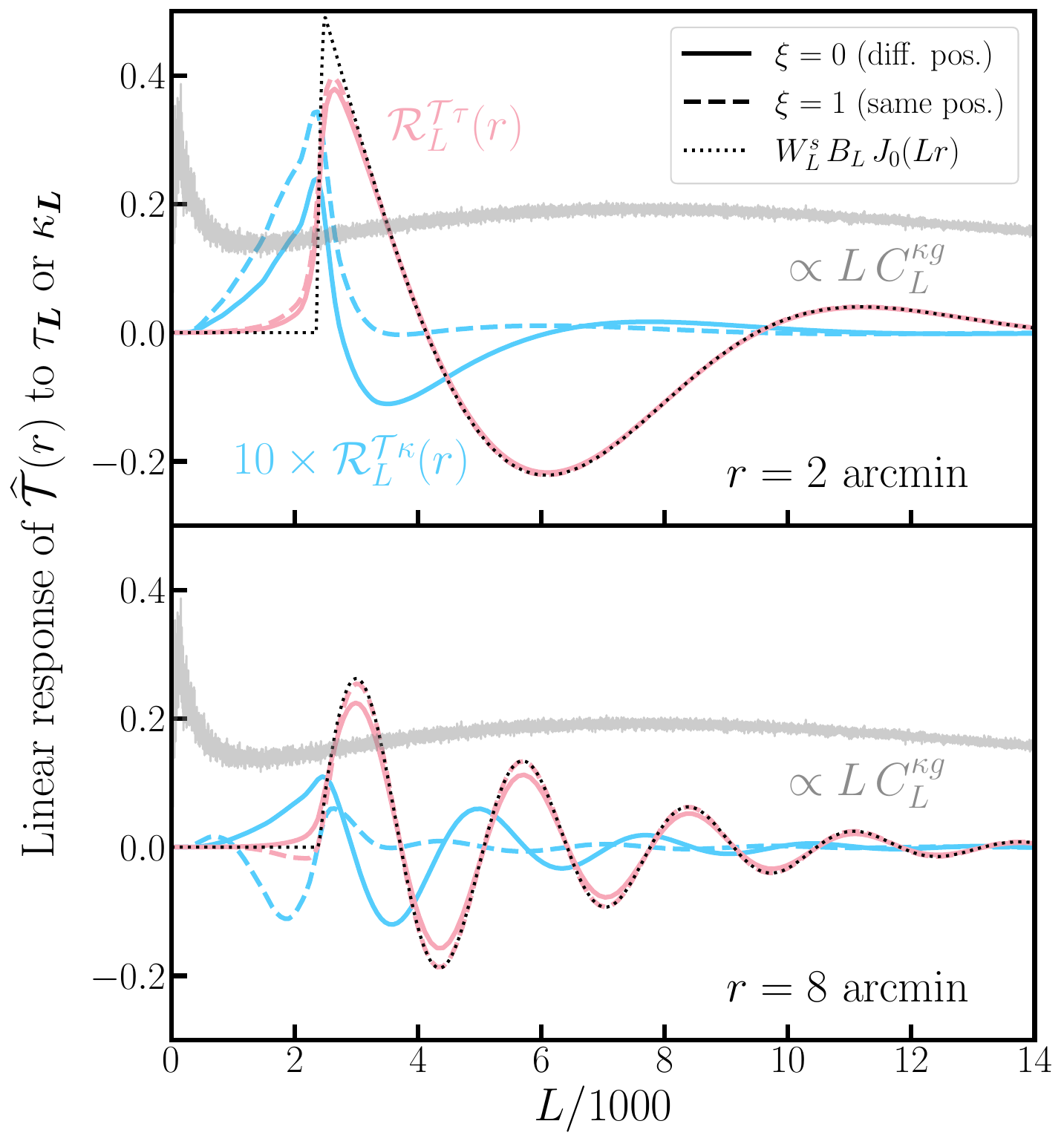}
    \caption{
    The isotropic response to lensing (Maya blue) and to patchy screening (Amaranth pink) for the experimental setup described in \S\ref{sec:sims}.
    In the top and bottom panels we fix $r=2$ and 8 arcmin respectively, while solid and dashed lines correspond to $T_l$ and $T_s$ being evaluated at offset or identical positions.
    Note that the response to lensing is multiplied by 10.
    In light gray we plot the CMB lensing cross-correlation measured from the \textsc{AbacusSummit} simulations for a DESI LRG-like set of mock galaxies \cite{boryana_companion_paper}.
    The dotted line shows the product of the high-pass filter, the beam $B_L$ and the Bessel function of the first kind $J_0(Lr)$.
    }
    \label{fig:rtk_rtt}
\end{figure}

For the case of an isotropic profile or when stacking with random orientations the response function simplifies to a function of two scalars $(L,r)$ rather than two vectors $(\bm{L},\bm{r})$. 
In Fig.~\ref{fig:rtk_rtt} we plot the isotropic response to lensing (Maya blue) and the isotropic response to patchy screening (Amaranth pink) for the analysis setup described in \S\ref{sec:sims}.
In the top and bottom panels we fix $r=2$ and 8 arcmin respectively, while solid and dashed lines correspond to $T_l$ and $T_s$ being evaluated at offset or identical positions in Eq.~\eqref{eq:stacked_estimator}. 

We emphasize that when $T_l$ and $T_s$ are evaluated at the same position the response to lensing (dashed blue) is confined to a smaller range of scales than when $T_l$ is evaluated at the galaxy center (solid blue), leaving the stacked estimator nearly unbiased on scales $L\gtrsim3000$ for $r\simeq2$ arcmin for the setup discussed in \S\ref{sec:sims}. 
We therefore recommend that when using a stacked estimator one evaluates $T_l$ and $T_s$ at the same position. 
The residual lensing bias can be subtracted following Eq.~\eqref{eq:stacked_hardening} or further suppressed (but not exactly nulled) by high-pass filtering the stacked map, which ``manually" sets $\mathcal{R}^{\mathcal{T} \tau}_L=\mathcal{R}^{\mathcal{T}\kappa}_L=0$ below some cutoff scale $L<L_c$.  

As is evident from Fig.~\ref{fig:rtk_rtt} the stacked response to patchy screening is very well approximated by the product of $W^s_L$ and the beam $B_L$ (see \S\ref{sec:sims} for a discussion of the beam) up to a geometric $J_0$ factor, implying that the stacked estimator reconstructs a high pass filtered $\tau(\bm{r})$ convolved with the beam to a very good approximation \cite{Schutt:2024zxe}.

\subsection{Mean hardening}
\label{sec:mean_hardening}

Field-level (\S\ref{sec:field_level_hardening}) and stacked hardening (\S\ref{sec:stacked_hardening}) require accurate small scale estimates of the lensing field. This can be seen from Fig.~\ref{fig:rtk_rtt}, where the response to lensing can be non-negligible for $L\lesssim8000$ (4000) when $T_l$ and $T_s$ are evaluated at offset (identical) positions. 
Faithfully reconstructing lensing on these scales with QEs is challenging due to contamination from extragalactic foregrounds \cite{2014ApJ...786...13V,2018PhRvD..97b3512F} and higher-order effects such as post-Born corrections \cite{Beck:2018wud}. 
In particular the ACT DR6 lensing map \cite{Frank_Ckk} is only reliable to $L_{\rm max}\sim3000$, while in principle one could push to $L_{\rm max}=6000$ for their baseline reconstruction scales ($\ell_{\rm max}=3000$).
Furthermore, on small scales the QE quickly becomes suboptimal.
While more optimal small-scale estimators have been proposed \citep[e.g.][]{2007NJPh....9..441H,Horowitz:2017iql,Millea:2021had,Bianchini:2022wte,2024JCAP...01..024S} these estimators are still subject to biases (primarily in temperature \cite{Qu:2024nqm}) from extragalactic foregrounds. 

In the absence of an accurate high resolution CMB lensing map one can instead estimate the lensing bias from an estimate of the large scale CMB lensing-galaxy cross-correlation, which we call \textit{mean hardening}. For simplicity we only consider the case of stacking with random orientations so that $\mathcal{R}^{\mathcal{T}\kappa}_{\bm{L}}(\bm{r})$ simplifies to a function of the scalars $(L,r)$. In Appendix~\ref{sec:general_stacked_linear_response} we show that the average lensing bias to the real space stacked estimator is
\begin{equation}
\label{eq:mean_hardening}
\boxed{
    \langle{\rm lensing\,\,bias\,}(r)\rangle = \int \frac{L\,dL}{2\pi} {\mathcal{R}}^{\mathcal{T}\kappa}_{L}(r) C^{\kappa g}_L,
}
\end{equation}
where $C^{\kappa g}_L$ is the CMB lensing-galaxy cross-correlation. 
Given an accurate measurement of $C^{\kappa g}_L$ on ``large scales" (e.g. $L\lesssim3000$) one can predict the small scale power from e.g. a power law extrapolation, halo model or hydrodynamic simulation.
While this approach comes with the cost of introducing a parametric model in the highly nonlinear regime, it has the benefit of being less sensitive to foreground contamination and higher order effects in the lensing reconstruction.

\section{Comparison with simulations}
\label{sec:sims}

We test both stacked (\S\ref{sec:stacked_hardening}) and mean (\S\ref{sec:mean_hardening}) hardening using a CMB lensing map and DESI LRG-like set of mock galaxies \citep[for details see][]{boryana_companion_paper} built from the \textsc{AbacusSummit} simulations \cite{2022MNRAS.509.2194H}. We adopt the following filters for the long and short maps respectively:
\begin{equation}
\label{eq:filters}
\begin{aligned}
    W^l_\ell 
    &=
    \begin{cases} 
        1 & \ell<2000 \\
        \cos\big(\frac{\pi}{300}(\ell-2000)\big) & {\rm otherwise} \\
        0 & \ell>2150
    \end{cases}
    \\
    W^s_\ell 
    &=
    \begin{cases} 
        0 & \ell<2350 \\
        \sin\big(\frac{\pi}{300}(\ell-2350)\big) & {\rm otherwise} \\
        1 & \ell>2500. 
    \end{cases}
\end{aligned}
\end{equation}
We assume that the short map is convolved with a Gaussian beam $B_\ell = \text{Exp}[-(\theta_{\rm FWHM}\ell)^2/16\ln 2]$ with a full width half max (FWHM) of 1.4 arcmin while the long map is assumed to be beam-deconvolved. To include the beam-convolution of the short map we substitute $W^s_\ell \to B_\ell W^s_\ell$ when computing the stacked response function (Eq.~\ref{eq:stacked_hardening}).

\begin{figure}[!h]
    \centering
    \includegraphics[width=\linewidth]{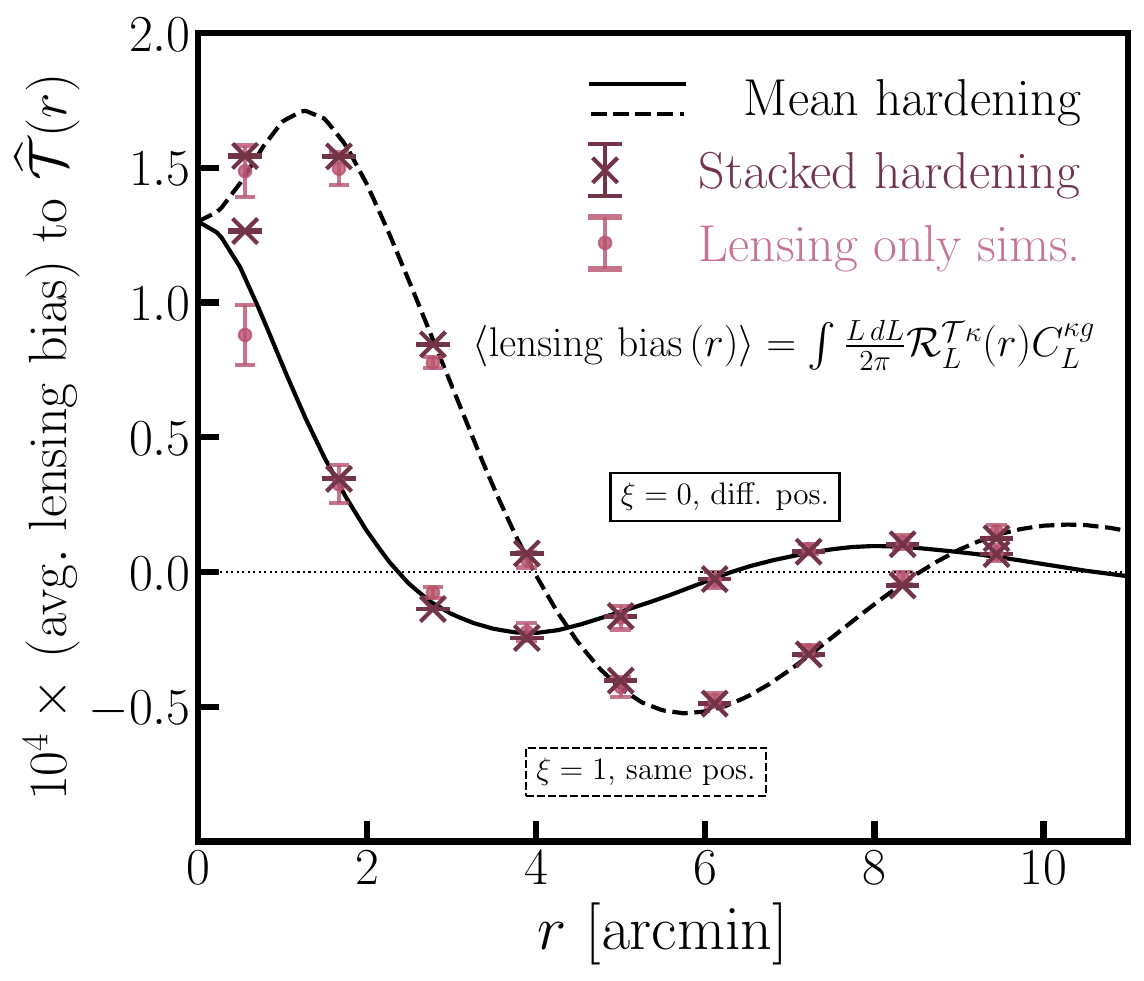}
    \caption{
    The predicted lensing bias from stacked (catawba) and mean (black lines) hardening for a DESI LRG-like set of mock galaxies built from the \textsc{AbacusSummit} simulations. 
    We find excellent agreement between both hardening approaches and the signal obtained when stacking on lensed CMB maps with no patchy screening signal (light catawba).
    Dashed and solid curves correspond to $T_l$ and $T_s$ being evaluated at identical or offset positions respectively.
    The profiles have been binned into 9 annuli whose 10 radial edges are linearly spaced between $r=0$ and $10$ arcmin.
    }
    \label{fig:avg_bias}
\end{figure}

In a companion paper \cite{boryana_companion_paper} we compute the profile obtained when lensed CMB maps with no patchy screening signal are stacked on LRG-like galaxies.
We plot the result in Fig.~\ref{fig:avg_bias} (light catawba) for the case where $T_l$ and $T_s$ are evaluated at offset ($\xi=0$, tracing the solid curve) or identical positions ($\xi=1$, tracing dashed). 
We use this measurement as a baseline from which to compare the predicted lensing bias from the stacked (catawba) and mean (black lines) hardening approaches.
We implement stacked hardening using the simulated convergence map while for mean hardening we use the cross-correlation measured from the simulation (plotted in light gray in Fig.~\ref{fig:rtk_rtt}).
With the exception of the first radial bin\footnote{
This is due to two effects: the bandpower window function has the largest impact on the comparison of the smooth to binned curves for the first radial bin, and in practice \cite{boryana_companion_paper} effectively applies a 0.5 arcmin smoothing to the galaxies when doing stacked hardening.
}
we find excellent agreement between all three approaches.

\section{Comments on ACT $\times$ unWISE}
\label{sec:unwise}

Refs.~\cite{Schutt:2024zxe,ACT:2024rue} proposed a modification to the stacked estimator, called the ``signed and thresholded" estimator, in which the long map $T_l$ appearing the numerator of Eq.~\eqref{eq:stacked_estimator} is replaced by its sign and cutouts with $|T_l(\bm{r}_i)|<40$ $\mu$K are discarded. This is equivalent to replacing $T_l$ with the weight $\mathcal{W}[T_l]\equiv
        \text{Sign}[T_l] H(|T_l|-T_c)
$,
where $H(x)$ is the Heaviside step function and $T_c=40$ $\mu$K. To ensure that $\widehat{\mathcal{T}}$ is correctly normalized one must also substitute $\langle T_l^2\rangle\to \langle \mathcal{W}[T_l]T_l\rangle$ in the denominator of Eq.~\eqref{eq:stacked_estimator}.
In Appendix \ref{sec:general_stacked_linear_response} we show that the linear response of this more general estimator to lensing (or any other extragalactic foreground) is identical to the $\mathcal{W}[T]=T$ case\footnote{Heuristically, this can be ``derived" by Taylor expanding \cite{taylor1715methodus} the weight function $\mathcal{W}[T]$.} provided that $\mathcal{W}[T]$ has a Fourier transform, which we compute for the signed and thresholded case in Appendix~\ref{app:fourier_transform}.
Thus, the leading order extragalactic foreground bias to the signed and thresholded estimator is identical to the standard stacked estimator (Eq.~\ref{eq:stacked_estimator}).

As an alternative to lens hardening (\S\ref{sec:lens_hardened_stacked_estimators}) one can suppress (but not precisely null) the lensing bias by filtering the stacked estimator to downweight modes with large response to lensing \citep[][see also \S\ref{sec:stacked_hardening}]{ACT:2024rue}. 
While this approach is formally suboptimal it has the advantage of being simpler to implement.  
We caution that since the clustering amplitude of a LSS tracer varies more slowly with the characteristic tracer mass than the mass itself, a lower-mass sample from e.g. LSST will presumably have a larger relative lensing bias than the more massive \textit{unWISE} galaxies. 
Thus, the precise filtering required for this approach to remain unbiased depends non-trivially on the galaxy sample considered, whereas lens hardening works ``out of the box" for any LSS tracer.

\begin{figure}[!h]
    \centering
    \includegraphics[width=\linewidth]{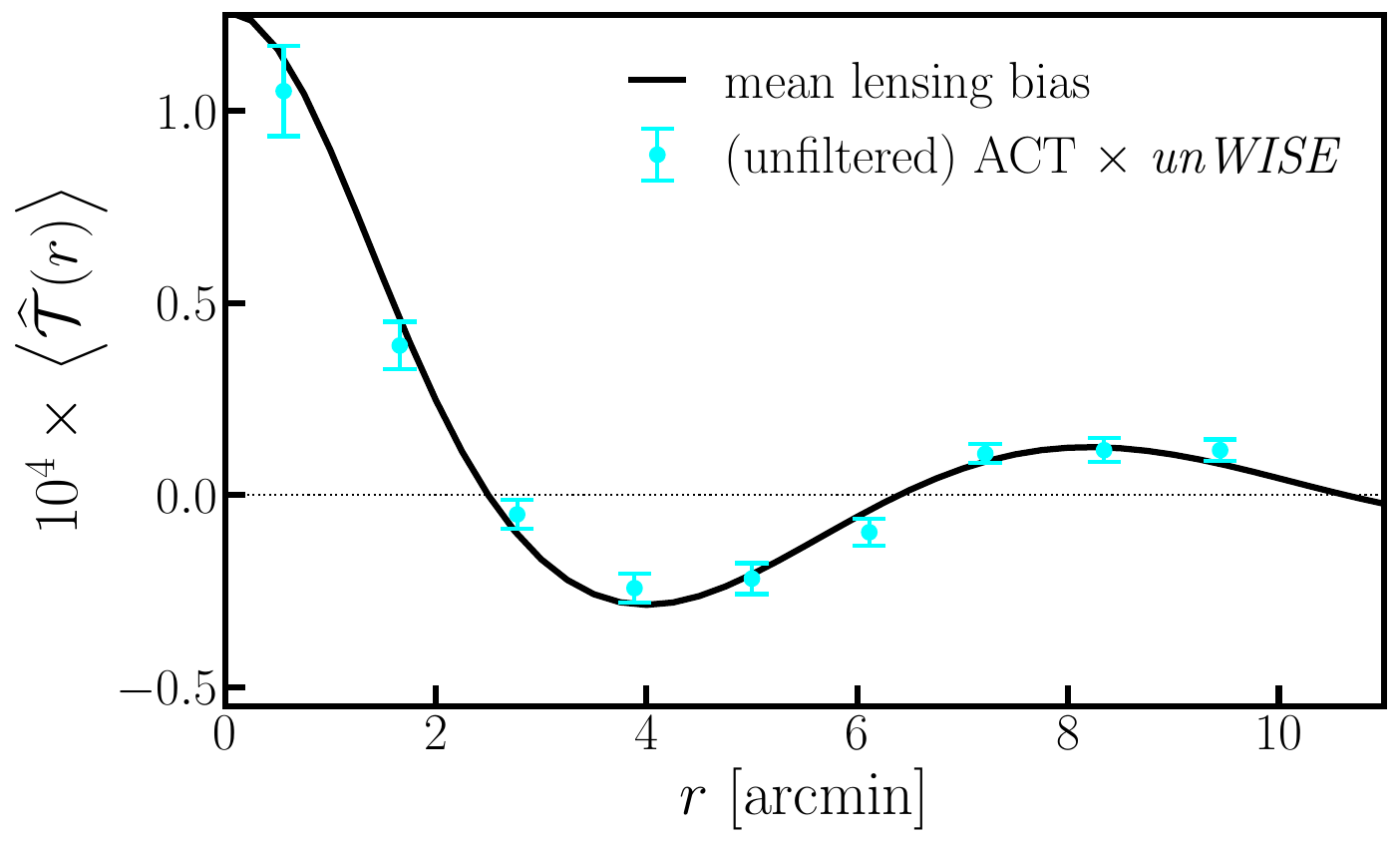}
    \caption{Mean lensing bias to the real space stacked estimator for the combined \textit{unWISE} blue and green sample (solid black line). In blue we show the unfiltered measurement from \cite{ACT:2024rue}.
    }
    \label{fig:unwise_bias}
\end{figure}

In Fig.~\ref{fig:unwise_bias} we show the (unfiltered) profile obtained from stacking ACT DR6 temperature maps on \textit{unWISE} blue and green galaxies \cite{Krolewski:2019yrv,2019ApJS..240...30S} using the long and short filters from \S\ref{sec:sims} \cite{ACT:2024rue}.
We note that the expected lensing bias (black) for the \textit{unWISE} sample dominates the signal if left unmitigated, which we estimate as follows.

We model the cross-correlation of the combined \textit{unWISE} green and blue sample with CMB lensing using the Kaiser-Limber approximation \cite{1953ApJ...117..134L,1992ApJ...388..272K,LoVerde:2008re}, a linear bias model and the non-linear matter power spectrum from the Aemulus $\nu$ simulations \cite{DeRose:2023dmk}. 
We use the fiducial redshift distributions from \cite{2024ApJ...966..157F} for the individual blue and green samples and add them to obtain the redshift distribution of the combined sample. 
We fix the cosmological parameters to their \textit{Planck} 2018 \cite{Planck:2018vyg} values and approximate the linear bias of the combined sample as the number-density weighted average of the fiducial blue and green linear biases from \cite{Krolewski:2021yqy}.
Likewise, we approximate the redshift evolution of the combined sample's magnification bias as the number-density weighted average of the (redshift independent) magnification biases measured in \cite{Krolewski:2019yrv} for the individual blue and green samples. 
The numerical implementation of the $C^{\kappa g}_L$ calculation is identical to that used in \cite{Kim:2024dmg,Sailer:2024coh}.
We assume that the short map is convolved with a 1.6 arcmin beam and evaluate $T_l$ at the galaxy center, corresponding to $\xi=0$ in Eq.~\eqref{eq:stacked_estimator}.

We find excellent agreement between the unfiltered profile and expected lensing bias, and hence no detection of patchy screening, despite not fitting any free parameters in our estimation of $C^{\kappa g}_L$. We subtract off the expected lensing bias in each radial bin (including the bandpower window) to obtain a debiased profile. We model the electron profile as a Gaussian with peak height $\tau_0$ and a FWHM of 2 arcmin that is then filtered by the response function ($\mathcal{R}^{\mathcal{T}\tau}$ with $\xi=0$, see Fig.~\ref{fig:rtk_rtt}).
When fit to the debiased profile we find $\tau_0<1.1\times10^{-4}$ at 68\% confidence. 



\section{Discussion and Conclusions}
\label{sec:discussion_and_conclusion}

Both stacked (\ref{sec:stacked_hardening}) and mean hardening (\ref{sec:mean_hardening}) are viable methods to mitigate the lensing induced bias with different assumptions made about the lensing field.
The former is more data driven and straightforward to adapt to oriented stacking and or optimal galaxy weighting, while the latter is more parametric and less sensitive to extragalactic foreground contamination. 
In \S\ref{sec:sims} we show that both approaches accurately reproduce the lensing bias found from simulated DESI LRG-like galaxies and lensed CMB maps.

Consistent with previous results \cite{2011arXiv1106.4313S}, we find that without mitigation the lensing contamination dominates the measured profile, making bias hardening essential for a robust screening measurement.
Applying mean hardening to previous measurements of ACT DR6 temperature maps stacked on \textit{unWISE} galaxies yields an upper bound on the screening signal consistent with expectations from recent kSZ measurements \citep[see the companion paper][]{boryana_companion_paper}.

While in this work we have focused on the leading order lensing bias, additional biases can arise from higher-order lensing effects and other extragalactic foregrounds. We leave a thorough investigation of these contaminants to future work, noting that in Appendix \S\ref{sec:general_stacked_linear_response} we show that the linear response of a generic class of stacked estimators (including the signed and thresholded estimator) to any extragalactic foreground is identical to that found for the ``standard" stacked estimator (Eq.~\ref{eq:stacked_estimator}).

For realistic surveys with partial sky coverage the stacked estimator is also biased by a ``mean field" which must be subtracted from Eq.~\eqref{eq:stacked_estimator} to obtain an unbiased profile. We compute the mean field bias in Appendix \ref{app:mean_field} and explicitly show that it can be estimated by stacking on random positions that are uniformly distributed within the galaxy survey mask. 
To obtain a precise estimate of the mean field relative to the measured profile one should ensure that the number of random positions is significantly larger than the number of galaxies used when stacking. 
An application of optimal lens hardening and mean field subtraction will be the subject of future work.

\section*{Acknowledgments}
We thank William Coulton and Emmanuel Schaan for useful discussions during the preparation of this manuscript. 
N.S. and S.F. are supported by Lawrence Berkeley National Laboratory and the Director, Office of Science, Office of High Energy Physics of the U.S. Department of Energy under Contract No.\ DE-AC02-05CH11231. B.H. thanks the Miller Institute for supporting her postdoctoral research.

\bibliographystyle{JHEP.bst}
\bibliography{main}

\onecolumngrid
\appendix

\section{Response functions of more general stacked estimators}
\label{sec:general_stacked_linear_response}

We define a more general stacked patchy screening estimator as weighted average of a filtered CMB map centered on a set of galaxy positions:
\begin{equation}
\label{eq:stacked_estimator_general}
    \widehat{\mathcal{T}}(\bm{r})
    \equiv
    -\frac{1}{N_g \langle \mathcal{W}[T_l]T_{l}\rangle}
    \sum_{i=1}^{N_g} \mathcal{W}[T_{l}(\xi\bm{r}+\bm{r}_i)] T_{s}(\bm{r}+\bm{r}_i),
\end{equation}
where $\bm{r}_i$ is the position of the $i$'th galaxy, $T_l(\bm{r}) = \int_{\bm{\ell}} e^{i\bm{\ell}\cdot\bm{r}} W^l_\ell T_{\bm{\ell}}$ is a filtered CMB map with isotropic (low pass) filter $W^l_\ell$ and likewise for $T_s(\bm{r})$, while $\mathcal{W}[T]$ is a general weighting function. As in the main text, $\xi$ is a number specifying where the long map should be evaluated relative to the short map.
Provided that the weighting function has a one-dimensional Fourier transform $\mathcal{W}_\omega$ we can rewrite 
\begin{equation}
    \widehat{\mathcal{T}}(\bm{r})
    =
    -\frac{1}{N_g \langle \mathcal{W}[T_l]T_{l}\rangle}
    \sum_{i=1}^{N_g} \int \frac{d\omega}{2\pi} \mathcal{W}_\omega e^{i\omega T_l(\xi\bm{r}+\bm{r}_i)}T_{s}(\bm{r}+\bm{r}_i).
\end{equation}
To compute the response of $\widehat{\mathcal{T}}(\bm{r})$ to lensing we average over primary CMB realizations in the presence of a fixed late time structure realization, which we denote as $\langle\widehat{\mathcal{T}}(\bm{r})\rangle'$. We Taylor expand the exponential and repeatedly Fourier transform to find
\begin{equation}
\label{eq:stacked_estimator_ft}
    \big\langle
    e^{i\omega T_{l}(\xi\bm{r}+\bm{r}_i)}
    T_{s}(\bm{r}+\bm{r}_i)
    \big\rangle'
    =
    \sum_{n=0}^\infty
    \frac{(i\omega)^{2n+1}}{(2n+1)!}
    \int_{\bm{\ell}\bm{\ell}_1\cdots\bm{\ell}_{2n+1}}
    e^{i\bm{\ell}\cdot(\bm{r}+\bm{r}_i)}
    e^{i(\bm{\ell}_1+\cdots+\bm{\ell}_{2n+1})\cdot(\xi\bm{r}+\bm{r}_i)}
    \langle 
    T^s_{\bm{\ell}} 
    T^l_{\bm{\ell}_1}
    \cdots 
    T^l_{\bm{\ell}_{2n+1}}
    \rangle',
\end{equation}
where we neglect terms with an odd number of $T$'s, as these terms vanish in the partial average. Note that for a fixed lensing realization the lensed CMB $T_{\bm{\ell}}$ is still a Gaussian random field, just not a statistically isotropic one, so we can use Wick's theorem to reduce 
$
\langle 
T_{\bm{\ell}} 
T_{\bm{\ell}_1}
\cdots 
T_{\bm{\ell}_{2n+1}}
\rangle'
$
to something like $\sim\langle T^2\rangle^{n+1}$. Note that only terms that are fully symmetric in the last $2n+1$ indices contribute to Eq.~\eqref{eq:stacked_estimator_ft}. The symmetric contribution to the $2n+2$ point function is given by
\begin{equation}
\begin{aligned}
    \langle 
    T_{\bm{\ell}} 
    T_{(\bm{\ell}_1}
    \cdots 
    T_{\bm{\ell}_{2n+1})}
    \rangle'
    &=
    (2n+1)
    \langle T_{\bm{\ell}} T_{(\bm{\ell}_1}\rangle' 
    \langle 
    T_{\bm{\ell}_2}
    \cdots 
    T_{\bm{\ell}_{2n+1})}
    \rangle'
    \\
    &= 
    \frac{(2n+1)!}{2^n n!} 
    \langle T_{\bm{\ell}} T_{(\bm{\ell}_1}\rangle' 
    \langle 
    T_{\bm{\ell}_2}
    T_{\bm{\ell}_3}
    \rangle'
    \cdots 
    \langle
    T_{\bm{\ell}_{2n}}
    T_{\bm{\ell}_{2n+1})}
    \rangle'.
\end{aligned}
\end{equation}
Substituting this in to Eq.~\eqref{eq:stacked_estimator_ft} and resumming gives
\begin{equation}
\begin{aligned}
\label{eq:neato_equation}
    \big\langle
    e^{i\omega T_{l}(\xi\bm{r}+\bm{r}_i)}
    T_{s}(\bm{r}+\bm{r}_i)
    \big\rangle'
    &=
    i\omega
    \int_{\bm{\ell}\bm{\ell}'}
    e^{i(\bm{\ell}+\bm{\ell}')\cdot\bm{r}_i}
    e^{i(\bm{\ell}+\xi\bm{\ell}')\cdot\bm{r}}
    \langle 
    T^s_{\bm{\ell}} T^l_{\bm{\ell}'} \rangle'
    \,\,\text{Exp}
    \bigg[
    -\frac{\omega^2}{2}
    \int_{\bm{\ell}\bm{\ell}'}
    e^{i(\bm{\ell}+\bm{\ell}')\cdot(\xi\bm{r}+\bm{r}_i)}
    \langle 
    T^l_{\bm{\ell}} T^l_{\bm{\ell}'} \rangle'
    \bigg].
\end{aligned}
\end{equation}
A nearly identical calculation can be carried out for the denominator: $\langle \mathcal{W}[T_l]T_l\rangle = \int d\omega \mathcal{W}_\omega \langle e^{i\omega T_l} T_l\rangle/2\pi$. To compute $\langle e^{i\omega T_l} T_l\rangle$, take Eq.~\eqref{eq:neato_equation} and replace $T^s$ with $T^l$, the partial average $\langle\cdots\rangle'$ with a full ensemble average $\langle \cdots\rangle$, and set $\xi=1$. The integrals inside and outside the exponential are identical and simplify to $\langle T^2_l\rangle$, giving
\begin{equation}
\begin{aligned}
    \langle \mathcal{W}[T_l(\bm{r})]T_{l}(\bm{r})\rangle
    =
    \langle T_l^2\rangle
    \int \frac{d\omega}{2\pi}
    e^{-\frac{\omega^2}{2} \langle T^2_l\rangle}
    i\omega \mathcal{W}_\omega.
\end{aligned}
\end{equation}
Putting this all together gives
\begin{equation}
\label{eq:partial_stack_average}
    \langle\widehat{\mathcal{T}}(\bm{r})\rangle'    
    =
    -\frac{1}{N_g\langle T^2_l\rangle}\sum_{i=1}^{N_g}
    \int_{\bm{\ell}\bm{\ell}'}
    e^{i(\bm{\ell}+\bm{\ell}')\cdot\bm{r}_i}
    e^{i(\bm{\ell}+\xi\bm{\ell}')\cdot\bm{r}}
    \langle 
    T^s_{\bm{\ell}} T^l_{\bm{\ell}'} \rangle'
    \frac{
    \int d\omega\,\omega\, \mathcal{W}_\omega 
    \,\,\text{Exp}
    \bigg[
    -\frac{\omega^2}{2}
    \int_{\bm{\ell}\bm{\ell}'}
    e^{i(\bm{\ell}+\bm{\ell}')\cdot(\xi\bm{r}+\bm{r}_i)}
    \langle 
    T^l_{\bm{\ell}} T^l_{\bm{\ell}'} \rangle'
    \bigg]
    }{
    \int d\omega\,\omega\, \mathcal{W}_\omega 
    \,\,\text{Exp}
    \bigg[
    -\frac{\omega^2}{2}
    \langle T_l^2\rangle
    \bigg]
    }.
\end{equation}

\subsection{Perturbative expansion}
To compute the response of $\widehat{\mathcal{T}}$ to lensing at a given order, we Taylor expand $\langle T_{\bm{\ell}} T_{\bm{L}-\bm{\ell}}\rangle'$ around a fixed lensing realization
\begin{equation}
    \langle T_{\bm{\ell}} T_{\bm{L}-\bm{\ell}}\rangle'
    =
    (2\pi)^2\delta^D_{\bm{L}} C^0_{\ell}
    +
    f^\kappa_{\bm{\ell},\bm{L}-\bm{\ell}}\,\kappa_{\bm{L}}
    +
    \sum_{n=1}^\infty
    \int_{\bm{\ell}_1\cdots\bm{\ell}_{n}}
    f^{\kappa^{n+1}}_{\bm{\ell},\bm{L}-\bm{\ell},\bm{\ell}_1,\cdots,\bm{\ell}_n}
    \,
    \Big(
    \kappa_{\bm{\ell}_1}
    \cdots
    \kappa_{\bm{\ell}_{n}}
    \kappa_{\bm{L}-\bm{\ell}_{1n}}\Big),
\end{equation}
and collect all terms in Eq.~\eqref{eq:partial_stack_average} at a given order in $\kappa$, where $f^{\kappa^{n+1}}_{\bm{\ell},\bm{L}-\bm{\ell},\bm{\ell}_1,\cdots,\bm{\ell}_n}$ defines the $n$'th order response function and we've adopted the shorthand $\bm{\ell}_{1n} \equiv \sum_{i=1}^n \bm{\ell}_i$. 
The higher order response functions ($f^{\kappa^n}$ for $n>1$) can be found by Taylor expanding the the lensed CMB in powers of the convergence
\begin{equation}
    T_{\bm{\ell}}
    =
    T^0_{\bm{\ell}}
    +
    \sum_{n=1}^\infty
    \int_{\bm{\ell}_1\cdots\bm{\ell}_n}
    K^{(n)}_{\bm{\ell},\bm{\ell}_1,\cdots,\bm{\ell}_n}
    \kappa_{\bm{\ell}_1}\cdots\kappa_{\bm{\ell}_n}
    T^0_{\bm{\ell}-\bm{\ell}_{1n}}
    \quad
    {\rm where}
    \quad
    K^{(n)}_{\bm{\ell},\bm{\ell}_1,\cdots,\bm{\ell}_n}
    =
    \frac{(-2)^n}{n!} \prod_{i=1}^n \frac{\bm{\ell}_i}{\ell_i^2}\cdot\big(\bm{\ell}-\bm{\ell}_{1n}\big).
\end{equation}
For example, the quadratic response function $f^{\kappa^2}$ is given by
\begin{equation}
\begin{aligned}
    f^{\kappa^2}_{\bm{\ell}_1,\bm{\ell}_2,\bm{\ell}_3}
    &=
    K^{(2)}_{\bm{\ell}_1,\bm{\ell}_1+\bm{\ell}_2-\bm{\ell}_3,\bm{\ell}_3} C^0_{\ell_2}
    +
    \frac{1}{2}
    K^{(1)}_{\bm{\ell}_1,\bm{\ell}_3}
    K^{(1)}_{\bm{\ell}_2,\bm{\ell}_1+\bm{\ell}_2-\bm{\ell}_3}
    C^{0}_{|\bm{\ell}_1-\bm{\ell}_3|}
    +
    (\bm{\ell}_1\leftrightarrow\bm{\ell}_2)
    \\
    f^{\kappa^2}_{\bm{\ell},\bm{L}-\bm{\ell},\bm{\ell}'}
    &=
    \frac{2\ell'_i(\bm{L}-\bm{\ell}')_j}{\ell'^2 |\bm{L}-\bm{\ell}'|^2}
    \bigg[
    \ell_i\ell_j
    C^0_\ell
    -
    (\bm{\ell}-\bm{\ell}')_i(\bm{\ell}-\bm{\ell}')_j
    C^0_{|\bm{\ell}-\bm{\ell}'|}
    \bigg]
    +
    (\bm{\ell}\leftrightarrow \bm{L}-\bm{\ell})
\end{aligned}
\end{equation}
where $C^0_\ell$ is the power spectrum of the unlensed CMB and summation over $i,j$ is implicit.

\begin{itemize}
\item\textbf{Linear order:} Expanding Eq.~\eqref{eq:partial_stack_average} up to linear order in $\kappa$ gives
\begin{equation}
\label{eq:first_order_lensing_bias}
\big\langle\widehat{\mathcal{T}}(\bm{r})\big\rangle'
=
\textcolor{gray}{\overline{\mathcal{T}}(r)}
+
\frac{1}{N_g}
\sum_{i=1}^{N_g}
\int_{\bm{L}}
e^{i\bm{L}\cdot\bm{r}_i}\,
\mathcal{R}^{\mathcal{T}\kappa}_{\bm{L}}(\bm{r})\,
\kappa_{\bm{L}}
\end{equation}
where
$
\overline{\mathcal{T}}(r)
=
-\big\langle T_l(\xi\bm{r}) T_s(\bm{r})\big\rangle/\langle T_l^2\rangle
$
is the (full-sky) mean field. 
The linear response function is
\begin{equation}
\mathcal{R}^{\mathcal{T}\kappa}_{\bm{L}}(\bm{r})
    =
    \frac{e^{i\xi\bm{L}\cdot\bm{r}}}{\langle T_l^2\rangle}
    \int_{\bm{\ell}}
    \Bigg[
    \hspace{2mm}
    \textcolor{gray}{
    \underbrace{\frac{\overline{\mathcal{T}}(r)}{2}
    \frac{d \ln\mathcal{Z}[T_l^{\rm rms}]}{d\ln T_l^{\rm rms}}
    W^l_{|\bm{L}-\bm{\ell}|}
    W^l_\ell}_{\text{vanishes in practice}}}
    -
    e^{i(1-\xi)(\bm{L}-\bm{\ell})\cdot\bm{r}}
    W^s_{|\bm{L}-\bm{\ell}|}
    W^l_\ell
    \Bigg]
    f^\kappa_{\bm{\ell},\bm{L}-\bm{\ell}},
\end{equation}
where we defined
$
\mathcal{Z}[T]
\equiv
\int d\omega
\,\omega \,\mathcal{W}_\omega\,
\text{Exp}[-(\omega T)^2/2]
$ and $T_l^{\rm rms}=\sqrt{\langle T_l^2\rangle}$.
In practice we assume the long and short filters are disjoint (i.e. $W^l_\ell W^s_\ell=0$) so that the full-sky mean field vanishes ($\overline{\mathcal{T}}=0$). 
In this case the response of $\widehat{\mathcal{T}}$ to lensing is independent of the weight function $\mathcal{W}[T]$. Note that this derivation is agnostic to the functional form of the lensing response function ($f^{\kappa}$). It immediately follows that the general stacked estimator's (Eq.~\ref{eq:stacked_estimator_general}) linear response to any extragalactic foreground is independent of the weighting function provided that the full-sky mean field vanishes.

Note that in principle the linear response is a function of the vectors $(\bm{r},\bm{L})$. This is relevant for oriented stacking, however, in this work we assume (uniform) random orientations, in which case the response can be replaced with its angle average, making it a function of the scalars $(r,L)$. 

We note that the linear order term can be isolated from simulations by lensing either the short or long map to linear order, leaving the remaining map unlensed, and averaging over the resulting two profiles.

\item\textbf{Quadratic order:} The quadratic contribution to Eq.~\eqref{eq:partial_stack_average} can be written as
\begin{equation}
    \big\langle\widehat{\mathcal{T}}(\bm{r})\big\rangle'
    \supset
    \frac{1}{N_g}
    \sum_{i=1}^{N_g}
    \int_{\bm{L}}
    e^{i\bm{L}\cdot\bm{r}_i}
    \int_{\bm{\ell}}
    \mathcal{R}^{\mathcal{T}\kappa^2}_{\bm{L},
    \bm{\ell}}(\bm{r})
    \,\,\kappa_{\bm{\ell}}\,
    \kappa_{\bm{L}-\bm{\ell}}
\end{equation}
where the quadratic response function is
\begin{equation}
    \mathcal{R}^{\mathcal{T}\kappa^2}_{\bm{L},
    \bm{\ell}'}
    (\bm{r})
    =
    \frac{e^{i\xi\bm{L}\cdot\bm{r}}}{\langle T_l^2\rangle}
    \int_{\bm{\ell}}
    \Bigg[
    \frac{1}{2}
    \frac{d\ln \mathcal{Z}[T_l^{\rm rms}]}{d\ln T_l^{\rm rms}}
    \mathcal{R}^{\mathcal{T}\kappa}_{\bm{\ell}'}(\bm{r})\,
    W^l_{|\bm{L}-\bm{\ell}'-\bm{\ell}|}\, W^l_{\ell}\,
    f^\kappa_{\bm{\ell},\bm{L}-\bm{\ell}'-\bm{\ell}}
    -
    e^{i(1-\xi)(\bm{L}-\bm{\ell})\cdot\bm{r}}\,
    W^s_{|\bm{L}-\bm{\ell}|}\,
    W^l_\ell
    \,f^{\kappa^2}_{\bm{\ell},\bm{L}-\bm{\ell},\bm{\ell}'}
    \Bigg],
\end{equation}
assuming that the mean field vanishes ($\overline{\mathcal{T}}=0$).
\end{itemize}





\section{Fourier transform of the signed and thresholded estimator}
\label{app:fourier_transform}

\begin{figure}[!h]
    \centering
    \includegraphics[width=0.5\linewidth]{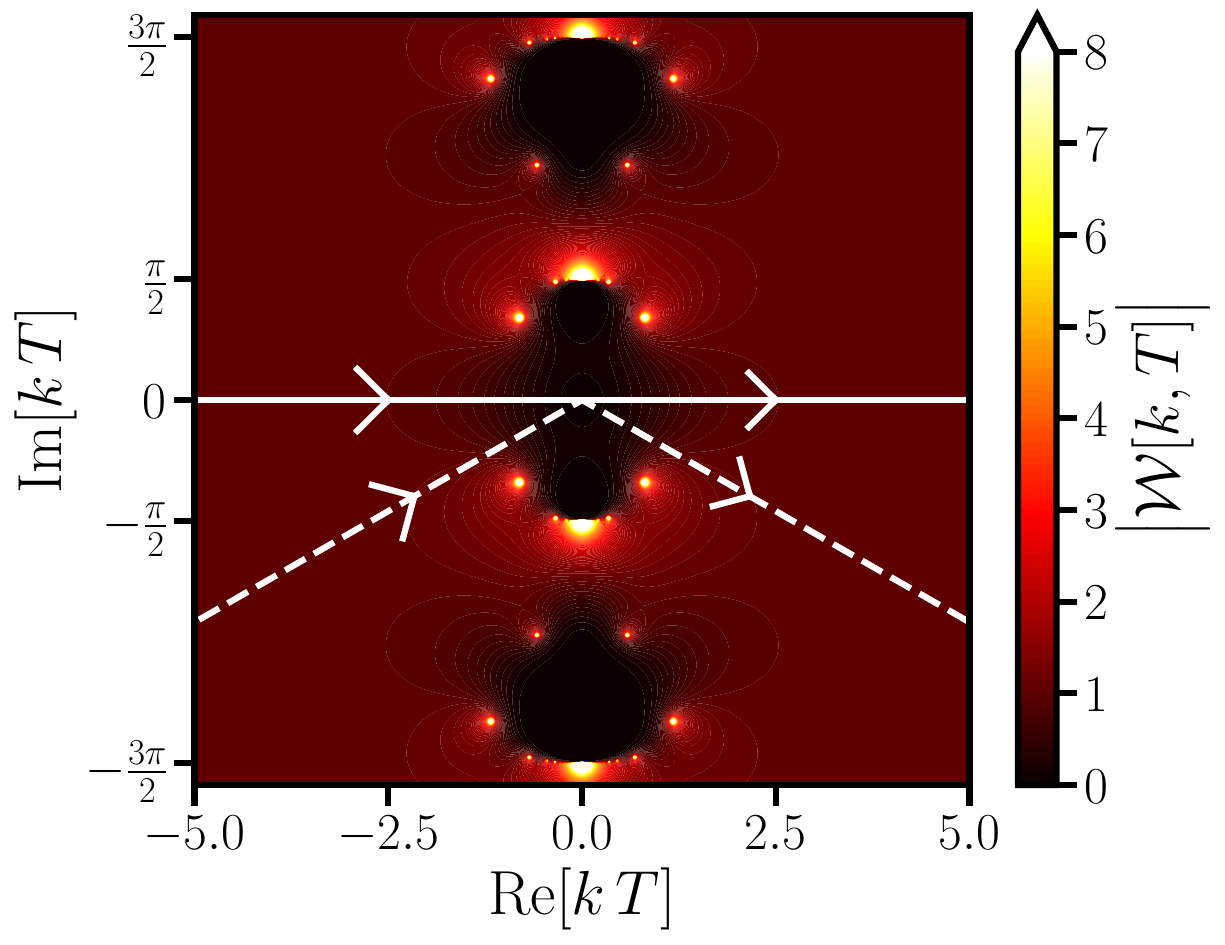}
    \caption{The pole structure of $\mathcal{W}[k,T]$ for $k\,T_c=0.5$.}
    \label{fig:contour_plot}
\end{figure}

Consider the signed and thresholded estimator
$
\mathcal{W}[T]=
        \text{Sign}[T] H(|T|-T_c)
$
where $H(x)$ is the Heaviside step function and $T_c$ is a cutoff temperature \cite{Schutt:2024zxe,ACT:2024rue}.
To compute the Fourier transform
$
    \mathcal{W}_\omega
    =
    \int_{-\infty}^\infty
    dT\,\,
    e^{-i\omega T}
    \,\mathcal{W}[T]
$
we first note that we may write $\mathcal{W}[T] = \lim_{k\to+\infty} \mathcal{W}[k,T]$ where 
\begin{equation}
    \mathcal{W}[k,T]
    =
    \frac{\tanh(k\, T)}{2}
    \bigg(
    1
    +
    \tanh\Big(k\,T\,\tanh(k\,T)-k\,T_c\Big)
    \bigg).
\end{equation}
The function $\mathcal{W}[k,T]$ is meromorphic in $T$ for any finite $k$. We plot $|\mathcal{W}[k,T]|$ in Fig.~\ref{fig:contour_plot} for $k\,T_c=0.5$. To ensure a convergent result for $\mathcal{W}_\omega$ we bend the contour slightly so that $T$ acquires a negative imaginary component as $|T|\to\infty$. This is illustrated in Fig.~\ref{fig:contour_plot}, where we take $T\to (1\pm i\epsilon )T$ to shift the contour from the solid to dashed white line. 
We are always free to make $\epsilon$ small enough to dodge any poles for a finite $k$, leaving us with (for $\omega>0$)
\begin{equation}
\begin{aligned}
    \mathcal{W}_\omega
    &=
    \lim_{k\to +\infty}
    \lim_{\epsilon\to 0^+}
    \left[
    \int_{T_c}^\infty dT 
    \,\,e^{-(i+\epsilon)\omega T}
    -
    \int_{-\infty}^{-T_c} dT 
    \,\,e^{-(i-\epsilon)\omega T}
    +\mathcal{O}(k^{-1})
    \right]
    \\
    &=
    \lim_{\epsilon\to 0^+}
    \frac{1}{\omega}
    \left[
    \frac{e^{-(i+\epsilon)\omega T_c}}{i+ \epsilon}
    +
    \frac{e^{(i-\epsilon) \omega T_c}}{i- \epsilon}
    \right]
    \\
    &=
    \frac{2}{i\omega}
    \cos(\omega T_c).
\end{aligned}
\end{equation}

\section{Mean field bias}
\label{app:mean_field}

Just as for CMB lensing, the stacked patchy screening estimator acquires a non-zero mean when the galaxy and CMB data have partial sky coverage (even if the long and short filters are disjoint). This ``mean field" must be subtracted from Eq.~\eqref{eq:stacked_estimator} to obtain an unbiased estimate of the gas profile. 
The mean field bias can be estimated by simply replacing the galaxy positions with a set of random positions uniformly dispersed within the galaxy mask. To rigorously show this, note that we can rewrite Eq.~\eqref{eq:stacked_estimator} as 
\begin{equation}
\begin{aligned}
\label{eq:stacked_estimator_rewrite}
    \widehat{\mathcal{T}}(\bm{r})
    &=
    \frac{-1}{\langle T^2_l\rangle}
    \int_{\bm{L}\bm{\ell}}
    \left[
    \frac{1}{N_g}
    \sum_{i=1}^{N_g}
    e^{i\bm{L}\cdot\bm{r}_i}
    \right]
    e^{i(\bm{L}-(1-\xi)\bm{\ell})\cdot\bm{r}}
    W^s_{|\bm{L}-\bm{\ell}|}
    W^l_\ell
    T_{\bm{\ell}} T_{\bm{L}-\bm{\ell}}.
\end{aligned}
\end{equation}
The sum over galaxy positions only includes positions within the galaxy mask $M(\bm{r})$, and is related to the masked density contrast $\tilde{g}$ via
\begin{equation}
\label{eq:masked_contrast_map}
    \tilde{g}(\bm{r})
    =
    \frac{A_M}{N_g}\sum_{i=1}^{N_g} \delta^D(\bm{r}-\bm{r}_i)
    -
    M(\bm{r}),
\end{equation}
where $A_M = \int d^2\bm{r} M(\bm{r})$ is the area within the mask $M$. Fourier transforming this expression and substituting it in to Eq.~\eqref{eq:stacked_estimator_rewrite} gives
\begin{equation}
\begin{aligned}
    \langle \widehat{\mathcal{T}}(\bm{r})\rangle
    =
    -\frac{1}{A_M \langle T^2_l\rangle}
    \int_{\bm{L}\bm{\ell}}
    e^{i(\bm{L}-(1-\xi)\bm{\ell})\cdot\bm{r}}
    W^s_{|\bm{L}-\bm{\ell}|}
    W^l_\ell
    \Big[
    \langle \tilde{g}_{-\bm{L}}
    T_{\bm{\ell}} T_{\bm{L}-\bm{\ell}}\rangle
    +
    M_{-\bm{L}}
    \langle T_{\bm{\ell}} T_{\bm{L}-\bm{\ell}}\rangle
    \Big]
    .
\end{aligned}
\end{equation}
The first term is the signal while the second term is the mean field bias. Even if the long and short filters are disjoint ($W_\ell^l W^s_\ell=0$), the mean field bias does not vanish provided that the galaxy \textit{and} CMB maps have partial sky coverage (in which case $M_{-\bm{L}}\langle T_{\bm{\ell}} T_{\bm{L}-\bm{\ell}}\rangle\neq 0$ for $L>0$). To isolate the second term one can choose a set of positions $\{\bm{r}_i\}$ from a distribution that is independent of the CMB data satisfying $\langle \tilde{g}\rangle=0$ (e.g. uniform-area sampling within the masked region).

\end{document}